\documentstyle{mn}
\input epsf.sty
\newif\ifAMStwofonts
\AMStwofontstrue
 
\def\Rin{R_{\rm in}}
\def\kT{k T_{\rm e}}

\def\taues{\tau_{\rm es}}

\def\Ne{N_{\rm e}}
\def\me{m_{\rm e}}
\def\vx{v_{\rm x}}
\def\vy{v_{\rm y}}
\def\vz{v_{\rm z}}

\ifoldfss
  \ifCUPmtlplainloaded \else
    \NewTextAlphabet{textbfit} {cmbxti10} {}
    \NewTextAlphabet{textbfss} {cmssbx10} {}
    \NewMathAlphabet{mathbfit} {cmbxti10} {} % for math mode
    \NewMathAlphabet{mathbfss} {cmssbx10} {} %  "   "    "
  \fi
  \ifAMStwofonts
    \ifCUPmtlplainloaded \else
      \NewSymbolFont{upmath} {eurm10}
      \NewSymbolFont{AMSa} {msam10}
      \NewMathSymbol{\upi}     {0}{upmath}{19}
      \NewMathSymbol{\umu}     {0}{upmath}{16}
      \NewMathSymbol{\upartial}{0}{upmath}{40}
      \NewMathSymbol{\leqslant}{3}{AMSa}{36}
      \NewMathSymbol{\geqslant}{3}{AMSa}{3E}

      \let\leq=\leqslant 
       \let\ge=\geqslant
    \fi
  \fi
\fi % End of OFSS
 
\ifnfssone
  \newmathalphabet{\mathit}
  \addtoversion{normal}{\mahit}{cmr}{m}{it}
  \addtoversion{bold}{\mathit}{cmr}{bx}{it}
  \newmathalphabet{\mathbfit} % math mode version of \textbfit{..}
  \addtoversion{normal}{\mathbfit}{cmr}{bx}{it} 
  \addtoversion{bold}{\mathbfit}{cmr}{bx}{it}
  \newmathalphabet{\mathbfss} % math mode version of \textbfss{..}
  \addtoversion{normal}{\mathbfss}{cmss}{bx}{n}
  \addtoversion{bold}{\mathbfss}{cmss}{bx}{n}
  \ifAMStwofonts
    \ifCUPmtlplainloaded \else
and
your
      \UseAMStwoboldmath
      \makeatletter
      \new@mathgroup\upmath@group
      \define@mathgroup\mv@normal\upmath@group{eur}{m}{n}
      \define@mathgroup\mv@bold\upmath@group{eur}{b}{n}
      \edef\UPM{\hexnumber\upmath@group}
      \new@mathgroup\amsa@group
      \define@mathgroup\mv@normal\amsa@group{msa}{m}{n}
      \define@mathgroup\mv@bold\amsa@group{msa}{m}{n}
      \edef\AMSa{\hexnumber\amsa@group}  
      \makeatother
      \mathchardef\upi="0\UPM19
      \mathchardef\umu="0\UPM16
      \mathchardef\upartial="0\UPM40
      \mathchardef\leqslant="3\AMSa36
      \mathchardef\geqslant="3\AMSa3E

      \let\leq=\leqslant 
       \let\ge=\geqslant 
    \fi
  \fi
\fi % End of NFSS release 1
   
\ifnfsstwo
  \DeclareMathAlphabet{\mathbfit}{OT1}{cmr}{bx}{it}
  \SetMathAlphabet\mathbfit{bold}{OT1}{cmr}{bx}{it}
  \DeclareMathAlphabet{\mathbfss}{OT1}{cmss}{bx}{n}
  \SetMathAlphabet\mathbfss{bold}{OT1}{cmss}{bx}{n}
  \ifAMStwofonts
    \ifCUPmtlplainloaded \else
      \DeclareSymbolFont{UPM}{U}{eur}{m}{n}
      \SetSymbolFont{UPM}{bold}{U}{eur}{b}{n}
      \DeclareSymbolFont{AMSa}{U}{msa}{m}{n}
      \DeclareMathSymbol{\upi}{0}{UPM}{"19}
      \DeclareMathSymbol{\umu}{0}{UPM}{"16}
      \DeclareMathSymbol{\upartial}{0}{UPM}{"40}
      \DeclareMathSymbol{\leqslant}{3}{AMSa}{"36}
      \DeclareMathSymbol{\geqslant}{3}{AMSa}{"3E}

      \let\leq=\leqslant 
       \let\ge=\geqslant
    \fi
  \fi  
\fi % End of NFSS release 2
      
\ifCUPmtlplainloaded \else
  \ifAMStwofonts \else % If no AMS fonts
    \def\upi{\pi}
    \def\umu{\mu}
    \def\upartial{\partial}
  \fi
\fi

\title{On the effect of coronal outflow on spectra formation in galactic black
 hole systems}
\author[ A. Janiuk, B. Czerny, P.T. \.Zycki]
       {A. Janiuk, B. Czerny, P.T. \.Zycki \\
        Nicolaus Copernicus Astronomical Center, Bartycka 18, 00-716 Warsaw, 
        Poland\\}
\begin{document}

\maketitle

\begin{abstract}
We present the results of both analytical and numerical calculations of 
the amplitude of the reflection component in X-ray spectra of galactic black 
hole systems. We take into account the anisotropy
of Compton scattering and the systematic relativistic bulk motion of the hot
 plasma. In the case of single scattering approximation the reflection from the disc surface is significantly enhanced due to the anisotropy of Compton scattering. On the other hand the
 calculations of multiple scattering  obtained using the Monte Carlo method 
show that the anisotropy effect is much weaker in that case. Therefore, the 
enhanced backscattered flux may affect the observed spectra only if the disc 
surface is highly ionized, which reduces the absorption in the energy band 
corresponding to the first Compton scattering.

\end{abstract}


\section{Introduction}

Hard X-ray spectra of the galactic black hole systems are well described by 
a power law primary emission along with the pronounced reflected component, 
which causes the observed flattening of the spectrum. 
The primary emission is likely to be produced by  Compton upscattering of soft
 photons on thermal electrons in hot, optically thin medium close to
a relatively cold accretion disc, being the source 
of seed photons for Comptonization (see e.g.\ review in Poutanen 1998). 
A fraction of the upscattered photons is directed towards the disk and
can be reflected from its surface, giving the rise to reflected continuum 
and fluorescent iron line emission (Lightman \& White 1988; George \& Fabian
1991).

Observational data for Cyg X-1 and other black hole systems 
in their hard/low state show often rather hard spectra (photon spectral index
 $\Gamma \sim 1.5-1.9$; Poutanen et al. 1997; Gierli\'nski et al. 1997; 
Dove et al. 1997; see also Poutanen 1998), while the amplitude 
of reflection $R$ covers the broad range of values between 0 and 2.
Moreover, $R$ and $\Gamma$ are correlated  (Zdziarski, Lubi\'{n}ski \& Smith 
1999; Revnivtsev, Gilfanov \&  Churazov 1999), in the sense that the
harder the spectrum, the smaller the amplitude of reflection.
The correlation exists both within the low/hard state and when sources
change their spectral state (\.{Z}ycki, Done \& Smith 1998).

These observations cannot be explained by the model in which static, 
continuous corona covers the cold accretion disc, as it predicts the 
power law slope $\Gamma \ge 2$ (i.e.\ rather soft spectra) and the reflection 
amplitude $R \sim 1.0$.
Among the possible models, which could reproduce the reflection amplitude in 
the range $R = 0 - 1$ there are two competitive: (i) cold disc disrupted in 
the inner part (e.g.\ Poutanen, Krolik \& Ryde 1997; 
Esin, McClintock \& Narayan 1997)
and (ii) a highly ionized, non-disrupted  disc 
(Nayakshin, Kazanas \& Kallman 2000; Ross, Fabian \& Young 1999).  
Detailed shape of the 
reflected continuum depends on the geometry, ionization state and abundances 
of elements in the scattering medium. However, spectral fitting does not 
always allow to constrain these parameters independently.
It is possible to explain the observed spectra of GBH in terms of 
weakly ionized, or neutral reflection from the disc which inner radius is of 
the order of 50 $R_{\rm g}$ 
(e.g. Done \& \.{Z}ycki 1999 for the hard state of Cyg X-1), as well as 
with highly ionized reflection from the disc 
extending to the marginally stable orbit (as suggested by Ross et
al.\ 1999 for the hard state of Cyg X-1; see also Done \& Nayakshin 2000).

The third possible model, in which both the reflection amplitudes  $R>1$
and $R<1$ are 
possible, is a mildly relativistic outflow/inflow in the corona 
(Beloborodov 1999). Relativistic aberration
reduces the hard X-ray flux scattered towards the disc, which leads to 
reduction of the reflected component and the soft flux from reprocessing 
entering the corona. 
In order to obtain quantitative agreement with observed spectral
indices and reflection amplitudes,
the model requires additional reduction of the soft flux intercepted by 
the hot plasma. This leads to the 'active regions'
geometry (e.g.\ magnetic flares above the disc; Haardt, Maraschi \& Ghisellini
1994) rather than a continuous corona. 
However, the coronal outflow may reproduce the reflection amplitude 
$R < 1$ and in order to obtain $R>1$ an inflow of the plasma 
must be postulated.

In this article we reanalyze the non-static corona model, with
plasma moving at relativistic speed in the direction 
perpendicular to the disc surface. We study the dependence of the amplitude of
 the 
reflection component on the bulk motion velocity, taking into account the 
thermal motion of electrons within the
plasma determined by the electron temperature $T_{\rm e}$ and we discuss the 
resulting shape of the spectra. We also check if high reflection 
amplitudes ($R>1$) might be explained in the frame of the outflow model 
but taking 
into account possible high ionization of the disc surface and the anisotropy 
of  Compton scattering.
We emphasize the importance of the first Compton scattering, which plays 
crucial role at lower energies. For highly ionized disc surface the effect of 
absorption by heavy elements is reduced and the reflected spectrum  makes 
a substantial contribution to the total spectrum in the $\sim 0.5-5$ keV 
band. 
Therefore we present firstly the semi-analytical calculations in the 
approximation of single scattering, and after that we perform numerical 
simulations of multiple scattering, which is responsible for the power law 
shape of the hard X-ray tail.

The contents of the paper is as following. In 
Section~\ref{sect:single} we analyze semi-analytically the amplitude of the 
reflection in a single
scattering approximation (after Ghisellini et al.\ 1991) but taking into 
account anisotropy of scattering within the rest frame of the electron, 
describing the 
thermal motion without the assumption of highly relativistic beaming and 
incorporating the systematic bulk motion (outflow) of the corona. 
Since the effect of multiple scattering is essential in a real situation, in 
Section~\ref{sect:multi} we repeat the computations for a corona of a given
optical depth and an electron temperature using a Monte Carlo method  for both
 outwards and inwards directions of bulk velocity.
The dependence of the reflection amplitude on the outflow velocity is presented in Section~\ref{sub_mc}. The spectra resulting from Monte Carlo computations are presented in Section~\ref{sub_rad}. The discussion of the results is 
given in Section~\ref{sect:diss}.

\section{Single scattering approximation for the amplitude of reflection}
\label{sect:single}

In this section we generalize the determination of the amplitude of the 
reflection
component derived by Ghisellini et al. (1991) for a slab geometry of a hot corona 
and anisotropic soft photon input from underlying  disc. Those results were obtained
assuming isotropic electron scattering in the rest frame of an electron and 
relativistic chaotic motion of electrons. We introduce subsequently the anisotropy
of the Thomson cross-section (Section~\ref{sub_thom}), we relax the assumption
that the thermal motion of the plasma is relativistic (Section~\ref{sub_therm}),
and finally we introduce the systematic bulk motion of the corona 
(Section~\ref{sub_bulk}).

We show, that including the 
effect of anisotropic soft photon distribution in Compton scattering process 
the reflection 
amplitude $R \ge 2$ can be obtained. It's value is somewhat larger when the angular dependence of Thomson cross-section is taken into account. On the other hand for low electron velocities ($\gamma < 2$) the angular distribution of scattered radiation becomes important and the value of $R$ is reduced. Adding the systematic coronal outflow reduces the anisotropy effect only in the range of high bulk velocities. 
 However all these results stand mostly for the 
first scattering, while for the multiple scattering in the cloud of 
optical depth $\tau \sim 1$ the reflection is weaker (see Section~\ref{sect:multi}).

\subsection{The effect of angular dependence of the Thomson cross-section}
\label{sub_thom}

In this section we calculate the power of Compton radiation scattered by
 electrons with isotropic relativistic velocity field $v=\beta c$, $\beta \sim 1$.
This approach is appropriate for a non-thermal plasma.  
The radiation field is anisotropic and 
incoming photons subtend a restricted solid angle. We assume the geometry is the 
same as in Ghisellini et al. (1991). However, in our calculations we take into 
account the angular dependence of Thomson cross-section (Rybicki \& Lightman, 1979). 
Therefore the emitted power is given by

%\begin{equation}
\begin{eqnarray}
\lefteqn{ P(\alpha,\gamma)={3 \over 8} {P_{\rm iso} \over (1+\beta^2/3)}\int_{\phi_{\rm min}}^{2\pi-\phi_{\rm min}} d\phi {}}
\nonumber \\
& & {} \int_{\theta_{\rm min}}^{\theta_{\rm max}}(1-\beta \cos\theta)^2 (1+\cos^2\theta)\sin \theta d\theta {}
\end{eqnarray}
%\end{equation}

Here $P_{\rm iso}$ is the power of isotropic emission and $P(\alpha)$ is the power of radiation 
scattered by electron of the angle $\alpha$ between its velocity and the axis of symmetry of 
incoming photons. The limits $\theta_{\rm min}$, $\theta_{\rm max}$ and $\phi_{\rm min}$are functions of 
$\alpha$ and are given by equations (see Ghisellini et al., 1991):

\begin{eqnarray} \label{eq:limits}
\theta_{\rm min}=\max(0, \alpha-\pi/2)
\nonumber \\
\theta_{\rm max}=\min(\alpha+\pi/2, \pi)
\nonumber \\
\end{eqnarray}

\begin{displaymath}
\phi_{\rm min} = \left\{ \begin{array}{ll}
0 & \textrm{for $0<\theta<\pi/2-\alpha$}\\ 
  & \textrm{or $3\pi/2-\alpha<0<\pi$}\\
\arccos[(\tan\theta\tan\alpha)^{-1}] & \textrm{otherwise}
\end{array} \right.
\end{displaymath}
%\end{eqnarray}

\begin{figure}
\epsfxsize = 100 mm 
\epsfbox[50 180 630 530]{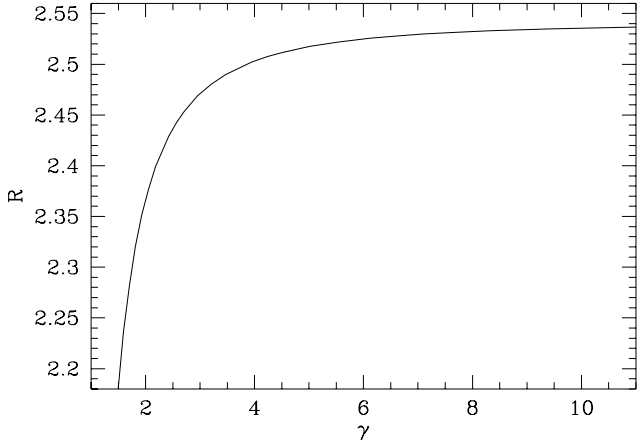}
\caption{The ratio of the total power emitted by electrons moving downwards to
 that of moving 
upwards as a function of electron's Lorentz factor $\gamma$.
\label{fig:ratio}}
\end{figure}

Here we follow the assumption of Ghisellini et. al (1991), that all photons 
are 
emitted in the direction of electron's velocity. This is justified in the 
case of 
$\gamma \gg 1$, when the direction of the motion of an electron (i.e the angle 
$\alpha$) can be identified with the viewing angle.
In Figure~\ref{fig:ratio} we plot the ratio of the total power emitted by 
electrons moving 
downwards to that of electrons moving upwards:
\begin{equation}
R(\gamma)={\int_{\pi/2}^{\pi}P(\alpha)\sin(\alpha)d\alpha \over 
\int_{0}^{\pi/2}P(\alpha)\sin(\alpha)d\alpha}
\end{equation}
It is worth seeing that this ratio saturates for large $\gamma$ at the value
 of $R \sim 2.55$, which is somewhat larger than that obtained under 
assumption of $d\sigma/d\Omega=\sigma_{\rm T}/4\pi$ ($R \sim 2.2$).
This result might be of possible importance for non-thermal Comptonization 
models. It seems however to be in conflict with the recent results of spectral
 fitting of the data for the soft state of Cyg X-1 (Gierli\'nski et al. 1999).
 
\begin{figure}
\epsfxsize = 100 mm 
\epsfbox[50 180 630 530]{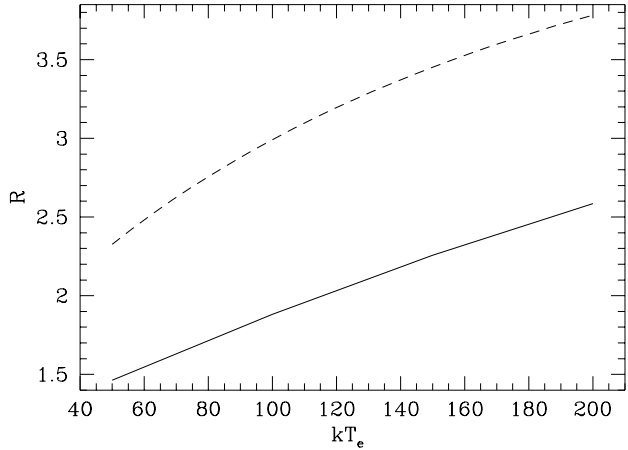}
\caption{The ratio of the total power emitted by electrons moving downwards to that of electron 
moving in one particular direction, as a function of electron temperature kT[keV], for an 
inclination angle $\alpha=30^\circ$, under the assumption of
efficient beaming (dashed line), and for the same inclination angle ($i=30^\circ$) but averaged over the whole range of electron's velocity direction (solid line).
\label{fig:temp}}
\end{figure}

In Figure~\ref{fig:temp} we plot the ratio of the power emitted downwards to 
that emitted 
by electron moving in one particular direction, given by
\begin{equation}
R(\alpha, kT)={\int_{\pi/2}^{\pi}P(\alpha)\sin(\alpha)d\alpha \over P(\alpha)}.
\end{equation}
This can be identified with the reflection amplitude seen by the observer with the viewing 
angle $\alpha$. We plot the result versus the electron temperature instead of
$\gamma$ in order to expose better the lower velocity part of the curve but we remind
that those results (dashed line) are not supposed to be accurate for not very highly 
relativistic velocities. We see that for low inclination angle the expected enhancement of
the reflection is very large, up to a factor 3.5 for typical plasma temperatures.

\subsection{The effect of mildly relativistic thermal motion}
\label{sub_therm}

If the plasma is rather thermal than non-thermal, the electron velocity is only
mildly relativistic and the assumption of all the emission being beamed in the
direction of motion of an electron used in Section~\ref{sub_thom} is not valid.
Therefore,  
in this section we consider the angular distribution of the scattered radiation, given by
\begin{equation}
P(\alpha,\gamma,\Theta_{\rm out})={P(\alpha,\gamma) \over 2\gamma^4(1+\beta \cos\Theta_{\rm out})^3}
\end{equation}
where $\Theta_{\rm out}=\pi$ means scattering in the direction of electron's movement. 
The accretion disc receives a fraction of scattered radiation that depends on the 
direction and velocity of electron
\begin{eqnarray}
\lefteqn{ P_{\rm disc}(\alpha, \gamma)= {}}
\nonumber\\
& & {}\int_{\Theta_{\rm out}^{\rm min}}^{\Theta_{\rm out}^{\rm max}} \int_{\phi_{\rm out}^{\rm min}}^{2\pi-\phi_{\rm out}^{\rm min}} P(\alpha,\gamma,\Theta_{\rm out}) \sin\Theta_{\rm out} d\Theta_{\rm out} d\phi_{\rm out}
\end{eqnarray} 
where $\Theta_{\rm out}^{\rm max}$, $\Theta_{\rm out}^{\rm min}$ and $\phi_{\rm out}^{\rm min}$ are given by equations (\ref{eq:limits}).

\begin{figure}
\epsfxsize = 100 mm 
\epsfbox[50 420 410 710]{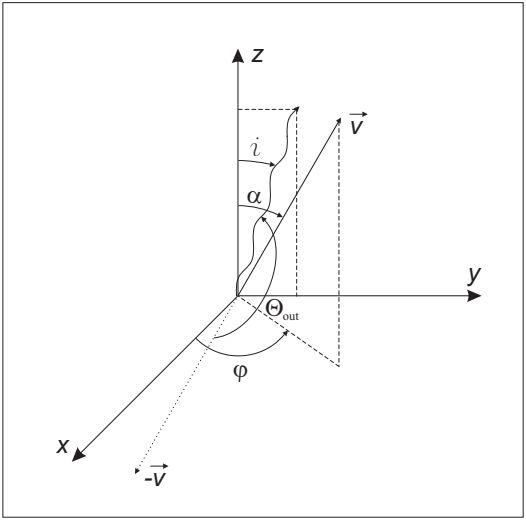}
\caption{The assumed geometry scheme. The electron is in the center of the 
reference frame and has the velocity $\vec v$ at the inclination $\alpha$ to 
the symmetry axis of the disc and at the angle $\varphi$ measured in the disc
 plane. The photon is emitted at the angle $i$ to the disc axis, and at the angle $(\Theta_{\rm out}-\pi)$ to the electron's velocity vector. 
\label{fig:uklad}}
\end{figure}

The viewing angle $i$ in this case cannot be identified with the direction of electron motion 
given by 
$\alpha$. Instead, we have a relation 
\begin{equation}
\cos\Theta_{\rm out}=-(\sin i \sin \varphi \sin \alpha+\cos i \cos \alpha).
\end{equation}
Here $\alpha$ and $\varphi$ determine the electron's velocity vector direction:
\begin{equation}
\vec v=v(\cos\varphi\sin\alpha,\sin\varphi\sin\alpha,\cos\alpha)
\end{equation}
In Figure~\ref{fig:uklad} we show the assumed geometry scheme. Note, that the
angle $\Theta_{\rm out}$ is measured from the vector $-\vec v$ to the direction
 of emitted photon, and therefore we keep the integration limits given by equations
(\ref{eq:limits})
 unchanged (see also Fig. 1 and Fig. 2 in Ghisellini et al. 1991).

\begin{figure}
\epsfxsize = 100 mm 
\epsfbox[50 180 630 530]{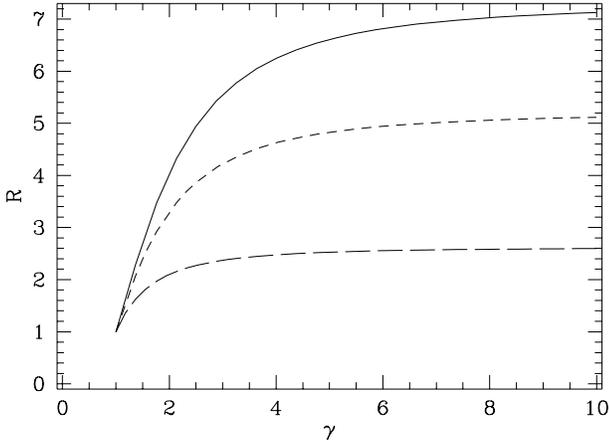}
\caption{The reflection amplitude for the viewing angle $i=0$ (solid line), $30^{\circ}$ 
(short-dashed line) and $60^{\circ}$ (long-dashed line) as a function of electron's Lorentz factor $\gamma$ in case of pure thermal motion.
\label{fig:refl}}
\end{figure}

We calculate the reflection amplitude averaged over the whole range of $\alpha$ and $\varphi$ as 
a function of electron velocity and viewing angle, according to the formula
\begin{equation}
R(\gamma,i)={1 \over 2\pi}{\int_{0}^{\pi} \int_{0}^{2\pi}P_{\rm disc}(\alpha, \gamma) \sin\alpha d\alpha d\varphi \over \int_{0}^{\pi}\int_{0}^{2\pi}P(\alpha,\gamma,\Theta_{\rm out}(\alpha,\varphi,i))\sin\alpha d\alpha d\varphi}.
\end{equation}
In  Figure~\ref{fig:refl} we plot the reflection amplitude as a function of $\gamma$ for three 
different values of viewing angle. The plot for an inclination $i=60^o$ roughly corresponds to the
value averaged over all inclinations, as presented in Figure~\ref{fig:ratio} for non-thermal 
plasma. We see that the anisotropy is only slightly reduced if the total collimation assumption is
relaxed. This effect is mostly seen in Figure~\ref{fig:temp} where we present with the solid 
curve low velocity (moderate
temperature) tail of the distribution.
 
%figure

\subsection{Mildly relativistic bulk motion}
\label{sub_bulk}

In this section we assume that bulk velocity vector is perpendicular to the disc surface and directed outwards. We calculate the 
net electron velocity as a sum of thermal chaotic motion and systematic (bulk) outflow. The angle between the net velocity 
vector and vertical axis, $\alpha'$, is connected with angle $\alpha$ via relativistic velocity 
transformation. Therefore the net Lorentz factor, $\gamma'$, depends on the angle $\alpha'$ as 
well as on bulk and thermal velocities. In this case the amount of reflection is given by:

\begin{eqnarray}
\lefteqn{ R(\beta_{\rm bulk},\beta_{\rm therm},i)={1 \over 2\pi} \times {} }
\nonumber\\
& & {}{\int_{0}^{\pi}\int_{0}^{2\pi}P_{\rm disc}(\alpha', \beta_{\rm bulk}, \beta_{\rm therm}) \sin\alpha' d\alpha' d\varphi' \over \int_{0}^{\pi}\int_{0}^{2\pi}P(\alpha',\beta_{\rm bulk},\beta_{\rm therm},\Theta_{\rm out}(\alpha',i))\sin\alpha' d\alpha' d\varphi'}.
\label{eq:bul}
\end{eqnarray}

\begin{figure}
\epsfxsize = 80 mm 
\epsfbox[50 140 520 700]{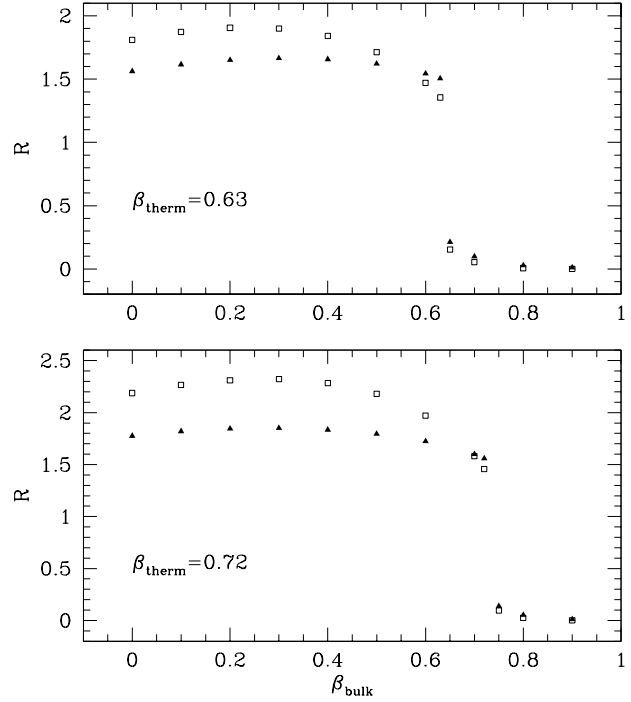}
\caption{The reflection amplitude for the viewing angle $i=0^{\circ}$ (boxes) and  $i=45^{\circ}$ (triangles) as a function of 
$\beta_{\rm bulk}$ for electron temperature $\beta_{\rm therm}=0.63$ (top panel) and  
$\beta_{\rm therm}=0.72$ 
(bottom panel).
\label{fig:bulk}}
\end{figure}

In  Figure~\ref{fig:bulk} we plot the dependence of the amount of reflection on the bulk velocity 
to the light velocity ratio, $\beta_{\rm bulk}$, for different values of electron temperature $\beta_{\rm therm}$ 
and viewing angle $i$. The electron temperature corresponds to a single value of electron velocity, as for velocity distribution the analytic approximation would not work. 

\begin{figure}
\epsfxsize = 100 mm 
\epsfbox[50 180 630 530]{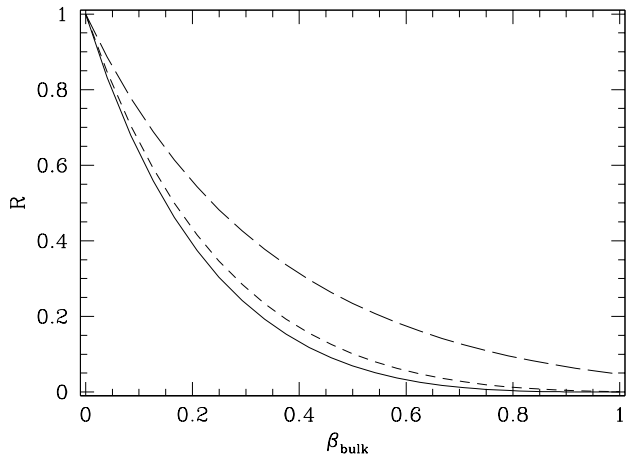}
\caption{The reflection amplitude for the viewing angle $i=0^{\circ}$ (solid line), $30^{\circ}$ (short-dashed line) and $60^{\circ}$ (long-dashed line) as a function of $\beta_{\rm bulk}$ for pure bulk motion.
\label{fig:beta}}
\end{figure}

In the case of $v_{\rm therm}=0$ we obtain the same solution as in Beloborodov (1999).
In Figure~\ref{fig:beta} we plot this
solution, calculated for different viewing angles.

The comparison of Figures~\ref{fig:beta} and \ref{fig:bulk} shows that single
scattering approach predicts very strong dependence of the amplitude of reflection
on the plasma temperature due to anisotropy of the Compton scattering. When the
thermal motions are important ($v_{\rm therm}/c \approx \beta_{\rm bulk}$) the first reflection is
significantly enhanced, by a factor of a two, and the higher the temperature the 
stronger the effect. The values of the amplitude of reflection cover the whole range
between 0 and $\sim 2$ for outflow solutions ($\beta_{\rm bulk} > 0$).

The effect of reflection enhancement drops rapidly when the bulk velocity is dominant ($\beta_{\rm bulk}> v_{\rm therm}/c$). In this case the maximum angle $\alpha'$ is much smaller than $\pi$ and the integration limits in the equation (~\ref{eq:bul}) must be changed. The reflection amplitude increases then with the value of assumed viewing angle $i$, while in the case of more significant thermal motion the trend is opposite.

\section{Reflected spectra from Monte Carlo simulations}
\label{sect:multi}

The roughly power law shape of the primary emission component in X-ray spectra
in  GBH is most likely due to the effect of multiple scatterings 
within the hot plasma.

It is well known from analytical solutions and Monte Carlo simulations 
that the contribution of the first scattering
to the total spectrum is rather specific (see Stern et al. 1995, 
Svensson 1996, Haardt, Maraschi \& Ghisellini 1997). It means, that when
 the hard X-ray spectrum forms via multiple 
scatterings, only the first one is influenced by the anisotropy of seed 
photon distribution.  This is the reason why any anisotropy effects are present in the low energy part of the spectrum.
In the thermal medium, even for small optical depths, the power law spectrum is shaped by multiple scatterings. Therefore
the semi-analytical computations, dealing with the first 
scattering process,  in the case of both  thermal and 
systematic bulk motion of the electrons within the corona can only serve as a 
guide and a help
to understand the numerical results.
 The fully reliable answer can only be provided by full
Monte Carlo simulations of the Comptonization process within the corona.

In this Section we compute the amplitude of the Compton reflected 
component using a Monte Carlo comptonization code and we show the 
corresponding spectra. We concentrate on highly ionized reflector, as the 
approximations used in Beloborodov (1999) may not be valid in that case. The 
code employs standard algorithms for
simulating the inverse Compton scattering, and it was written following 
descriptions by
Pozdnyakov, Sobol \& Sunyaev (1983) and G\'{o}recki \& Wilczewski (1984).
Modifications to the code necessary to implement the bulk motion are
described in Appendix~A.
Following Beloborodov (1999) we assume that the 
comptonizing region as a whole is stationary and its geometry is that 
of a slab.

\subsection{Reflection amplitude}
\label{sub_mc}

\begin{figure}
\epsfxsize = 100 mm 
\epsfbox[30 180 720 680]{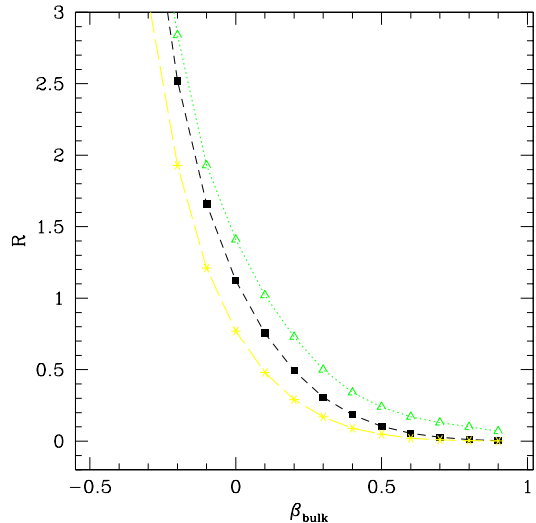}
\caption{
 Amplitude of the Compton-reflected component as a function of the bulk
 outflow velocity, $\beta_{\rm bulk}$, computed using the Monte Carlo method for
 multiple scattering in the plasma of optical depth  $\taues=0.8$ and 
electron temperature
 $kT_{\rm e}=100$ keV. Open triangles indicate the results for the inclination 
angle $\cos i=0.3$, stars mark the results for $\cos i=0.9$, and solid squares
result from the angle averaged spectrum.
 \label{fig:mcarlo}
}
\end{figure}

The bulk velocity vector is assumed perpendicular to the plane of the disc
but the direction of the velocity can be both outwards and towards the disc.
The disc is a source of soft photons for the comptonization. Photons 
backscattered from the cloud form radiation illuminating the disc. The
usual Compton reflection process is then simulated by another Monte Carlo
routine (\.{Z}ycki \& Czerny 1994), assuming
the abundances given in Morrison \& McCammon (1983). The opacities were 
computed using the code described in  Done et al.\ (1992) for the ionization 
parameter $\xi \equiv F_{\rm X}/(n_{\rm e} r^2) = 10^{4}$. 
Further scattering of the 
reflected photons in the hot comptonizing cloud is not considered, 
since the intercepted fraction
would be geometry-dependent (e.g.\ factor $\mu_{\rm s}$ in Beloborodov 1999).

The reflection amplitude $R$ is defined here (cf.\ Beloborodov 1999 and
Section~\ref{sub_bulk}), 
as the ratio of the energy integrated fluxes:
\begin{equation}
R(\beta_{\rm bulk}) = {F_{\rm back}(\beta_{\rm bulk}) \over F_{\rm dir}
(\beta_{\rm bulk}, i)}.
\end{equation}
Here $F_{\rm dir}$ is the flux  directly  escaping from the comptonizing cloud 
towards an observer, at the inclination angle $i$, 
and $F_{\rm back}$ is the flux directed towards the disc.
We note that our definition of $R$ would only be equivalent to that
used in spectral fitting, if the shapes of spectra used in models
fit to the data were exactly the same as in our simulations (for {\it both\/} 
the primary and the reflected components). Since this is not necessarily 
the case in practice, translating an amplitude $R$ inferred from spectral
modelling to a bulk velocity may not be accurate.

The parameters of the comptonizing cloud: the electron temperature $\kT$
and optical depth $\taues$ are chosen so that the resulting comptonized 
spectra correspond to typical spectra of GBH.
Figure~\ref{fig:mcarlo} shows the resulting amplitude of the reprocessed
component as a function of the plasma bulk velocity $\beta_{\rm bulk}$. 

Comparison of the Figures~\ref{fig:mcarlo} and \ref{fig:bulk} shows that the
enhancement of the reflection due to the anisotropy of the first scattering 
is sharply reduced if the subsequent scatterings are taken into account. 
Only for rather high inclination angles is the obtained reflection amplitude 
larger than 1.0 for $\beta_{\rm bulk}>0$, as the larger the viewing angle, 
the fewer  photons are able to escape from the slab towards the observer.
This  means,
in particular, that bulk velocities directed {\it towards\/} the disc are still
necessary in some cases in order to explain full range of the $\Gamma$--$R$ 
relation found by Zdziarski et al. (1999), as the
enhancement of flux towards the disc due to plasma bulk motion is required to 
explain $R>1$ seen in some sources.

\subsection{Radiation spectra of outflowing corona}
\label{sub_rad}

\begin{figure}
\epsfxsize = 100 mm 
\epsfbox[30 180 720 680]{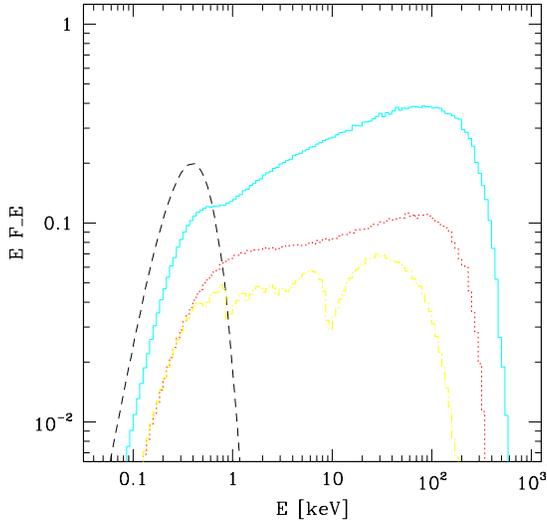}
\caption {
 Comptonized spectra resulting from the
 outflowing corona model obtained by Monte Carlo simulations. 
Parameters were chosen so that
 the spectra correspond to typical spectra of low/hard state of GBH:  
 $\kT=100$ keV, $\taues=0.8$.
 The bulk velocity is $\beta_{\rm bulk}=0.3$. The dashed line represents the 
soft photon input, solid line shows the comptonized continuum scattered 
towards the observer (angle averaged), the short-dashed line shows the
 backscattered spectrum and the short-long-dashed line represents the 
reflected spectrum for ionization parameter, $\xi = 10^4$.
 \label{fig:MCspecs}
}
\end{figure}

In Figure~\ref{fig:MCspecs} we present an example of the overall spectrum 
calculated from the model of the outflowing corona. We choose the following 
values for the model
parameters:
$\kT=100$ keV, $\tau=0.8$, $k T_0 = 0.1$ keV and the bulk velocity
$\beta_{\rm bulk}=0.3$. Adopted soft photon temperature corresponds to a typical value 
for galactic black holes. Such a parameterization is convenient if we do not
consider the energy balance within the corona. We assume that the corona
is a continuous medium, i.e. we neglect the clumpiness of the corona described
by the parameter $\mu_{\rm s}$ in the model of Beloborodov (1999) since we 
also  neglect the
secondary reprocessing of the reflected component through the corona.

The continuous line shows the radiation emitted towards an 
observer (averaged over the entire hemisphere, roughly corresponding to
an inclination of 60$^\circ$) while the short- dashed line shows
the component backscattered towards the disk. Since the first scattering 
dominates soft X-ray band for galactic sources, the backscattered radiation
in this band is enhanced, as predicted by analytical results presented in 
Section~\ref{sect:single}. However, hard X-ray part is dominated by multiple
scattered photons, anisotropy effect is smeared off and the backscattered 
component is not enhanced in this band. The net effect is therefore the 
systematic difference between the spectral slope of the back-scattered 
radiation and forward-scattered radiation. This effect was discussed for 
a corona without a bulk motion in
a number of papers (e.g. Stern et al. 1995). Therefore, the 
continuum formed in the corona and emitted towards an observer is slightly 
curved, particularly in the soft X-ray band, instead of being a simple power 
law with a high energy cut-off, as frequently assumed in spectral analysis 
of the
data. 

The backscattered continuum is subsequently reflected by the disk surface 
which in our calculations is assumed to be ionized ($\xi=10^4$).
We show this spectral component in the Figure~\ref{fig:MCspecs}.

The reflected component in the outflowing corona model is again partially
reprocessed by the corona. The effect depends on the corona clumpiness since
only a fraction of radiation $\mu_{\rm s}$ (following the notation of Beloborodov, 
1999) would pass again through the hot plasma. In the present paper we
neglect this secondary reprocessing since it is essential only if
$\mu_s$ is close to 1 and the optical depth is close to 1. However, in detail 
modeling this effect should be rather taken into account.

\section{Discussion}
\label{sect:diss}

 It is generally assumed that in  galactic X-ray sources and active galactic
 nuclei
soft X-ray radiation originates from the geometrically thin and optically 
thick accretion disc while the hard flux is produced by Comptonization in 
optically thin plasma outside the disc. Part of the hard X-ray flux that is 
directed towards the observer is detected in the form of power law continuum 
and the part of the flux scattered back to the disc surface produces so called
 'reflection hump' in the spectrum over 10 keV (Lightman \& White 1988;
Pounds et al.\ 1990; Done et al.\ 1992) as well as the iron $K_{\alpha}$ line 
near 6.4 keV (\.Zycki \& Czerny 1994).

The observed values of the amplitude of the reflection component cover 
broader range  than initially expected ($R \sim 1$ for AGN in Pounds et 
al.\ 1990). 
For some X-ray sources the observed reflection seems to be weak, which was
modelled either by the disruption of the inner disc and X-ray irradiation 
of its 
outer parts, or by the high ionization state of illuminated medium, 
which results
 in steepening of the  reflected continuum and makes it indistinguishable from
 the primary power law (Ross et al. 1999). 

Both interpretations can explain the observed correlation between $R$ and 
$\Gamma$. In the model with a disrupted cold disc partially overlapping 
the innermost hot flow the observed correlation is due to radiative coupling 
of the two components, as the amount of overlap varies (Zdziarski et al.\  
1999). In the model with strongly ionized 
disc surface the correlation is governed by the variable optical depth of the 
ionized scattering layer (Nayakshin et al.\ 2000), which determines the 
effective  albedo of the disc and the soft flux from thermalized fraction 
of  the illuminating X-rays. Both models can explain the reflection 
amplitude values $R\leq 1$.

However, the strength of reflection was for some sources found to be $R>1$, 
which may either indicate that reflecting medium subtends a  solid angle 
larger than $2\pi$, or that the radiation directed towards the disc is 
enhanced. The latter is possible when the scattering process is anisotropic 
or may be due to  the velocity of systematic bulk motion directed towards
the disc.

In this article we show that the anisotropy effect is important mostly for the
 first scattering and  weakens in numerical Monte Carlo simulations performed 
for multiple scattering in the plasma parameterized by optical depth and 
electron temperature. The maximum amplitude reached in the case of highly 
ionized reflector is only $R \sim 1.1$. Therefore, the systematic 
mildly-relativistic motion towards the disc, or the 'coronal inflow', is 
required to explain the higher values of $R$. On the other hand, in order to 
produce $R>1$ in the disrupted disc model it would be necessary to allow
for reflection from outer regions of a thickened disc
(e.g. Shakura \& Sunyaev 1973) and/or from the dusty/molecular torus
around central black holes of AGN.
Since the absorption of irradiating X-rays below $\sim 5$ keV is reduced 
for a highly ionized disc surface,  the reflected photons may contribute to 
the soft X-ray excesses observed in the spectra of GBH in the hard/low
state (Cyg X-1; Di Salvo et al.\ 1999, and in preparation). 
We show that this contribution is higher when the  radiation backscattered 
to the disc is enhanced as a result of the anisotropy  of the first 
scattering.   

The hard to low state transition characteristic for many accreting black
hole systems is in the 'disc plus sphere' model connected with the change 
of the inner radius of the cold disc. In the hard spectral state the cold 
disc is pushed outward while the hot inner plasma is responsible for hard 
X-ray radiation. In the soft state the thermal disc emission dominates as the 
cold disc extends almost to the marginally stable orbit. However, the physical
mechanism of such behaviour is currently unclear. In the outflowing corona
model the outflow velocity is the basic control parameter. The hard spectral
state would then correspond to rapid
coronal expansion, while the  spectrum with dominating soft component would 
be produced during the vertical collapse of the hot gas. Again the physical
mechanism of such dependence is unclear, and in particular it is unclear
how the required changes of either $\Rin$ or $\beta_{\rm bulk}$ could be driven
by changing accretion rate.

In conclusion, the basic predictions of all the proposed models are similar 
and only detailed computations of spectral and temporal behaviour and 
comparison with the high 
quality data, as expected from Chandra and XMM, may allow for a distinction 
between them. 

\section*{Acknowledgments}

This work was supported in part 
by grant 2P03D01816, 2P03D01718 and 2P03D01519
 of the Polish State Committee for 
Scientific Research.

\appendix

\section{The Monte Carlo code}

Our initial comptonization code was written following closely  descriptions
given by Pozdnyakov, Sobol \& Sunyaev (1983) and G\'{o}recki \& Wilczewski 
(1984), and applied to model X-ray spectra of accretion disc with accreting
advective corona by Janiuk, \. Zycki \& Czerny (2000). 
In order to include the effect of bulk motion of the plasma,
we needed to make two major modifications.

Firstly, the average scattering cross section has to be modified.
The escape probability of a photon is given by
\begin{equation}
P(d) = \exp\left( -\int_0^d \Ne \langle\sigma\rangle d l\right),
\end{equation}
where $\Ne = \int N(\bmath{v}) d^3 v$ is the electron density, $N(\bmath{v})$
is the electron velocity distribution, $d$ is the distance to the cloud
boundary along the direction of photon motion, $\bmath{\Omega}$, and
\begin{equation}
\label{equ:smean}
\langle\sigma\rangle = {1\over \Ne} \int N(\bmath{v}) (1-\bmath{v}\cdot 
\bmath{\Omega}/c) \sigma(x) d^3 v
\end{equation}
is the scattering cross section averaged over $N(\bmath{v})$. Here 
\begin{equation}
x = 2 { h \nu \over \me c^2} \gamma (1-\bmath{v}\cdot \bmath{\Omega}/c)
\end{equation}
is the energy of an incoming photon in the electron rest frame,
$\gamma$ is the Lorentz factor and $\sigma(x)$ is the Klein-Nishina cross
section. For an isotropic $N(\bmath{v})$ e.g.\ Maxwell distribution, 
$\langle\sigma\rangle$ is a function
of photon energy only (for a given $\kT$).

With the non-zero bulk velocity we cannot use the method presented by
Pozdnyakov et al.\ (1983) to evaluate the integral in 
Eq.~(\ref{equ:smean}), since $N(\bmath{v})$ is no longer isotropic.
The presence of the specific direction -- the bulk velocity vector 
$\bmath{\beta}$ -- 
introduces an additional angular dependence of $\langle\sigma\rangle$.
We now have to compute the integral as in Eq.~(\ref{equ:smean}) but
with $\bmath{v}$ in the dot-product $\bmath{v}\cdot\bmath{\Omega}$
replaced by total electron velocity, $\bmath{u}$. Here $\bmath{u}$ is
the sum (in the sense of Lorentz transformation) of the thermal velocity
and the bulk velocity.
We compute this 3-D integral numerically. Introducing a coordinate system
with the $z$-axis along the bulk velocity and the $x$-axis along the
direction of photon motion we obtain $\bmath{\Omega} = 
(\sin\theta,0,\cos\theta)$, where $\theta$ is the angle between 
$\bmath{\Omega}$ and $\bmath{\beta}$. Applying the Lorentz transformation 
we obtain the  electron velocity in the disc frame,
\begin{equation}
\bmath{u} = \left( {\vx \over \gamma\left(1+\beta {\vz\over c}\right)},
                   {\vy \over \gamma\left(1+\beta {\vz\over c}\right)},
            { \vz + \beta c \over 1+\beta {\vz\over c}} \right),
\end{equation}
with $\gamma = (1-\beta^2)^{-1/2}$, and $\vx,\ \vy\ {\rm and}\ \vz$ -- 
components of the electron's thermal velocity.
This enables to compute $\bmath{u}\cdot\bmath{\Omega}$ and calculate
the required integral, which is now additionally a function of photon's 
direction of motion, $\theta$. 
We used the procedure {\tt quad3d} from Press et al. (1992) to evaluate
the integral.

The second modification concerns the scattering event. Since this is modeled
in the electron rest frame, we introduced a pair of additional
Lorentz transformations of a photon's momentum: from
the disc/corona frame to the frame comoving with the bulk velocity before
simulating the scattering event, and the reverse transformation after
the scattering event.

\ \\
This paper has been processed by the authors using the Blackwell
Scientific Publications \LaTeX\  style file.


\bigskip

\bigskip

\begin{thebibliography}{}
\bibitem []{} Beloborodov A.M., 1999, ApJ, 510, L123
\bibitem []{} Di Salvo T., Done C., \.Zycki P.T., Burderi L., Robba N.R., 1999,
 Astrophysical Letters and Communications, 38, 261
\bibitem []{} Done C., Mulchaey J.S., Mushotzky R.F., Arnaud K.A., 1992, ApJ, 395, 275
\bibitem []{} Done C., \.{Z}ycki P. T. 1999, MNRAS, 305, 457
\bibitem []{} Done C., Nayakshin S. 2000, ApJ, submitted
\bibitem []{} Dove J.B., Wilms J., Maisack M., Begelman M.C., 1997, ApJ, 487, 759
%\bibitem []{} Ebisawa K., et al., 1994, PASJ, 46, 375
\bibitem []{} Esin A. A., McClintock J. E., Narayan R., 1997, ApJ, 489, 865
%\bibitem []{} Fabian A.C., Rees M.J., Stella L., White N.E., 1989, MNRAS, 238, 729 
\bibitem []{} George I.M., Fabian A.C., 1991, MNRAS, 249, 352
\bibitem []{} Ghisellini G., George I.M., Fabian A.C., Done C., 1991, MNRAS, 248, 14
\bibitem []{} Gierli\'nski M., Zdziarski A. A., Done C., Johnson W. N.,
 Ebisawa K., Ueda Y., Haardt F., Phlips B. F., 1997, MNRAS, 288, 958
\bibitem []{} Gierli\'nski M., Zdziarski A. A., Poutanen J., Coppi P. S.,
 Ebisawa K., Johnson W. N., 1999, MNRAS, 309, 496
\bibitem []{} G\'orecki A., Wilczewski W., 1984, Acta Astron., 34, 141
\bibitem []{} Haardt F., Maraschi L., Ghisellini G., 1994, ApJ, 432, L95
\bibitem []{} Haardt F., Maraschi L., Ghisellini G., 1997, ApJ, 476, 620
\bibitem []{} Janiuk, A., \. Zycki, P.T., Czerny, B., 2000, MNRAS (in press)
\bibitem []{} Lightman A.P., White T.R., 1988, ApJ, 335, 57
\bibitem []{} Morrison R., McCammon D., 1983, ApJ, 270, 119
%\bibitem []{} Nandra K., George I.M., Mushotzky R.F., Turner T.J., Yaqoob T., 1997, ApJ, 488, L91
\bibitem []{} Nayakshin S., Kazanas D., Kallman T., 2000, ApJ, in press,
(astro-ph/9909359)
\bibitem []{} Pounds K.A., Nandra K., Steward G.C., George I.M., Fabian A.C., 1990, Nat., 344, 132
\bibitem []{} Poutanen J., Krolik J.H., Ryde F., 1997, MNRAS, 292, L21
\bibitem []{} Poutanen J., 1998, in Abramowicz M.A., Bj\"ornsson G., Pringle J.E.. eds, Theory of Black Hole Accretion Discs, Cambridge Univ. Press, p. 100
\bibitem []{} Pozdnyakov L.A., Sobol I.M., Sunyaev R.A., 1983, ASPR, 2, 189
\bibitem []{} Press W.H., Teukolsky S.A., Vetterling W.T., Flannery B.P. 
    1992, ``Numerical Recipes in Fortran'', CUP
\bibitem []{} Revnivtsev M., Gilfanov M., Churazov E., 1999, A\&A Letters,
submitted, (astro-ph/9910423)
\bibitem []{} Rybicki G.B., Lightman A.P., 1979, ``Radiative Processes in Astrophysics'', Wiley, New York
\bibitem []{} Ross R.R., Fabian A.C., Young A.J., 1999, MNRAS, 306, 461
\bibitem []{} Shakura N.I., Sunyaev R.A., 1973, A\&A, 24, 337
\bibitem []{} Stern, B.E., Poutanen, J., Svensson, R., Sikora, M., Begelman, M.C., 1995, ApJ, 449, L13
\bibitem []{} Svensson, R., 1996, ApJS, 120, 475
%\bibitem []{} Turner T.J., George I.M., Nandra K., 1998, ApJ, 508, 648
\bibitem []{} Zdziarski A.A., Lubi\'nski P., Smith D.A., 1999, MNRAS, 303, L11
\bibitem []{} \.Zycki P.T., Czerny B., 1994, MNRAS, 266, 653
\bibitem []{} \.Zycki P.T., Done C., Smith D., 1998, ApJ, 496, L25


\end{thebibliography}
\end{document}